\documentclass[showpacs,preprintnumbers,amsmath,amssymb]{revtex4}

\usepackage{graphicx}
\usepackage{dcolumn}
\usepackage{bm}

\newcommand{\simgt}{\lower.5ex\hbox{$\; \buildrel > \over \sim \;$}}
\newcommand{\simlt}{\lower.5ex\hbox{$\; \buildrel < \over \sim \;$}}

\begin{document}
\preprint{~}
\title{
Constraint on the cosmological $f(R)$ model from the multipole 
power spectrum of the  SDSS LRG sample
and prospects for a future redshift survey}%
\author{Kazuhiro {Yamamoto}$^1$}
\email[Email:]{kazuhiro@hiroshima-u.ac.jp}
\author{Gen Nakamura$^1$}
\email[Email:]{gen@theo.phys.sci.hiroshima-u.ac.jp}
\author{Gert H\"{u}tsi$^{2}$}
\email[Email:]{gert@aai.ee} 
\author{Tatsuya Narikawa$^1$}
\email[Email:]{narikawa@theo.phys.sci.hiroshima-u.ac.jp}
\author{Takahiro Sato$^1$}
\email[Email:]{sato@theo.phys.sci.hiroshima-u.ac.jp}

\affiliation{
$^{1}$Department of Physical Science, Hiroshima University,
Higashi-Hiroshima 739-8526,~Japan
\\
$^{2}$Tartu Observatory, EE-61602 T\~{o}revere, Estonia}

\date{\today}

\begin{abstract}
A constraint on the viable $f(R)$ model is investigated by 
confronting theoretical predictions with the multipole power 
spectrum of the luminous 
red galaxy sample of the Sloan Digital Sky survey data release 7.
We obtain a constraint on the Compton wavelength parameter 
of the $f(R)$ model on the scales of cosmological large-scale structure. 
A prospect of constraining the Compton wavelength parameter 
with a future redshift survey is also investigated.
The usefulness of the redshift-space distortion
for testing the gravity theory on cosmological scales 
is demonstrated.
\end{abstract}

\maketitle

\def\bfk{{\bf k}}
\section{Introduction}
Experimental tests of gravity on the scale of the Solar 
System show good agreement with predictions 
of general relativity (e.g., \cite{Adelberger}).
The nature of the Newtonian gravity is the attractive force, which 
naturally predicts a decelerated expansion of the universe. 
Contrary to this expectation, it has been discovered 
that our universe is undergoing an accelerated expansion epoch 
\cite{Riess98,Perlmutter99,WMAP03}. 
Though the accelerated expansion is explained by introducing
a cosmological constant, its small but nonzero value
cannot be explained naturally \cite{Weinberg}.
The problem might be deeply rooted in the nature of fundamental 
physics. 

This problem has attracted many researchers, and many works have been 
done, both theoretically and observationally. 
As a generalisation of the cosmological constant, a dynamical 
field, called the dark energy model and its variants, are proposed 
to explain the accelerated expansion of the universe (see \cite{Peebles} 
and references therein). 
As an alternative to the dark energy model, modification of 
gravity may explain the accelerated expansion. 
General relativity is not considered to be the complete theory,
because its quantum theory cannot be formulated in a well defined manner.  
The theory of gravity might need to be reformulated within a more general 
framework. 

From the observational point of view, the constraint on the gravity 
theory on cosmological scales has not been well investigated, 
compared with the constraint on the scales of the Solar System.
Many future projects to produce large galaxy surveys are in progress 
or planned \cite{DETF06,sdss3,sumire,lsst,ska,Robberto}, 
which aim to explore the nature of the dark energy. 
These surveys are useful for testing the theory of gravity at 
cosmological scales (e.g., \cite{Yamamoto06}).
The dynamical dark energy models may have similar expansion rates as models 
of modified gravity, but predict different histories for the growth of 
structures.
The key to testing the gravity theory is the measurement of the 
evolution of cosmological perturbations, 
as many authors have concluded recently  
\cite{Uzan,Shirata,Sealfon,
Linder05,Ishak06,UzanII,
CahnLinder,Yamamoto07,Huterer07,Kunz07,Spergel08,Song09}.

The cosmic microwave background anisotropies are useful for investigating 
the cosmological perturbations through the measurements of the 
integrated Sachs-Wolfe effect or the lensing effect on the 
angular power spectrum \cite{Smoot}.
Imaging surveys of galaxies are also useful through the weak 
lensing statistics or cluster number counts \cite{JainTakada,Bean}. 
Similarly, redshift surveys of galaxies are helpful for testing gravity 
\cite{Linder,Guzzo,Yamamoto08,WhitePercival,Peacock09,Reyes}. 
In the present paper, we revisit the problem of testing the gravity 
theory through a measurement of the multipole power spectra in
the sloan digital sky survey (SDSS) luminous red galaxy (LRG)
sample \cite{Yamamoto08}. Measuring
the multipole power spectra  is a way to estimate the redshift-space 
distortions, which reflects the linear growth rate of the matter 
density perturbations \cite{Cole,Hamilton,YBN}. 

Many authors have investigated the clustering nature of the 
SDSS LRG sample \cite{Eisenstein,Hutsi,PercivalI,PercivalII,
Tegmark,PercivalIII,Okumura,HutsiII,Cabre,Sanchez}.
In the references \cite{Percival2009,Reid2009}, recent results on 
LRGs from the SDSS data release (DR) 7 are reported.
In the reference \cite{Cabre}, a test of gravity is considered
using the observed anisotropic correlation function.
Three of the authors of the present paper have shown that
the SDSS LRG sample is useful to test the gravity 
theory by measuring the quadrupole power spectrum of 
galaxy distribution, which represents the redshift-space 
distortions \cite{Yamamoto08}. 
In the present paper, we revisit the issue of testing the
gravity theories on the cosmological scales using the SDSS 
LRG sample of the DR 7, especially focusing 
on the $f(R)$ gravity model.

The $f(R)$ models proposed in \cite{HuSawicki,Starobinsky,Tsujikawa,Nojiri}
are viable models of modified gravity, which include some function of 
the Ricci scalar, $f(R)$, added to the Einstein Hilbert action.
As the modification of gravity
involves the introduction of extra degree of freedom in general, 
one must be careful with the resulting behaviour. Furthermore, any
theory must reduce to the general relativity on the 
scales of the Solar System. 
In the $f(R)$ model, the general relativity is supposed 
to be recovered by the {\it chameleon mechanism} \cite{KhouryWeltman,Mota}, 
which hides the field of the extra degree of freedom because 
the mass of the field becomes large for a dense region.
The cosmological bounds on the $f(R)$ model have been investigated 
with the cosmic microwave background anisotropies \cite{SPH}
and also using the abundance of galaxy clusters \cite{Schmidt}. 
However, our approach is based on the redshift-space 
distortion \footnote{After we have completed this manuscript, 
we noticed of the paper by Girones et al.~\cite{Girones}, in which 
the similar cosmological constraint on the $f(R)$ model is investigated.}.

This paper is organised as follows: In section 2, we
briefly review the $f(R)$ model and the characteristic
evolution of the matter density perturbation. 
In section 3, we present our results for the multipole power 
spectrum of the SDSS  LRG sample of the DR 7. 
In section 4, cosmological constraint is discussed 
by confronting the observed multipole spectra with the theoretical 
predictions. In section 5, a prospect of constraining
the $f(R)$ model is discussed on the basis of the Fisher 
matrix analysis, assuming a future large redshift survey.
Section 6 is devoted to summary and conclusions. Throughout 
this paper, we use units in which the velocity of light equals 1,
and adopt the Hubble parameter $H_0=100h {\rm km/s/Mpc}$ with $h=0.7$.

\section{$f(R)$ gravity model}
In this section, we briefly review the $f(R)$ model,
proposed in the references \cite{HuSawicki,Starobinsky,Tsujikawa,
Nojiri}.
In general, higher derivative terms are expected in the low energy 
effective action of gravity. Inspired by this, the $f(R)$ model 
introduces some function of the Ricci scalar $f(R)$, adding to the
Einstein Hilbert action. We consider the theory defined by
\begin{eqnarray}
  S={1\over 16\pi G} \int d^4 x \sqrt{-g} \left(R+f(R)\right) 
+S_{\rm m},
\end{eqnarray}
where $S_{\rm m}$ is the action of the matter. 
Many aspects of the $f(R)$ model have been investigated; 
see e.g. \cite{Sotiriou,Felice} for a review (cf., \cite{Carloni,Kobayashi}).
We assume that the chameleon 
mechanism is responsible for the recovery
of the general relativity on the Solar-System scales.
The chameleon mechanism is a nonlinear effect.
Recently, the effect on the quasi-nonlinear power spectrum
is investigated based on the perturbative approach or the 
numerical simulations \cite{Koyama09,Oyaizu08a,Oyaizu08b,Schmidt09a}.
This nonlinear chameleon effect becomes influential
in the nonlinear regime. In the present paper, however, we can 
neglect the nonlinear chameleon effect because we need to consider 
only rather large scales, $k\simlt0.2h{\rm Mpc}^{-1}$.

For the viable model, the function $f(R)$ must satisfy 
some conditions. We consider the model where the asymptotic
form of $f(R)$ can be expressed by
\begin{eqnarray}
  f(R)\simeq -2 \Lambda\left[1-\left({R_c\over R}\right)^{2n}\right],
\label{viablefr}
\end{eqnarray}
where $\Lambda$ is the cosmological constant, $n$ is a constant
that specifies the $f(R)$ model, and $R_c$ is also a 
constant with the same dimension as that of the Ricci scalar.
The background expansion of this $f(R)$ model is 
well approximated by that of the $\Lambda$CDM model.

It is known that the additional term $f(R)$ involves the introduction
of an extra degree of freedom. Namely, $f_R\equiv df/dR$ corresponds to the 
extra degree of freedom, which behaves like a scalar field. 
From the above action, one can derive the equation for $f_R$,
\begin{eqnarray}
 \nabla_\mu\nabla^\mu f_R={1\over 3}(R+2f-Rf_R)+{8\pi G\over 3}(-\rho+3P),
\label{eqoffr}
\end{eqnarray}
where $\rho$ and $P$ are the energy density and the pressure of the matter, 
respectively. 
If we regard the right hand side of equation (\ref{eqoffr})
as the derivative of the effective potential, $dV_{\rm eff}/df_R$, 
the mass of $f_R$ can be read 
\begin{eqnarray}
m^2={d^2 V_{\rm eff}\over d f_R^2}={1\over 3}
\left({1+f_R\over f_{RR}}-R\right).
\end{eqnarray}
The viable $f(R)$ theory satisfies $f\ll R$, and $|f_R|\ll 1$.
Assuming $Rf_{RR}\ll 1$, the mass of the extra degree of freedom is
\begin{eqnarray}
m^2\simeq {1\over 3}{1\over f_{RR}},
\end{eqnarray}
where $f_{RR}=d^2 f/dR^2$.
Thus, $f_{RR}>0$ is required to avoid the extra degree of freedom 
to become tachyonic. This extra degree of freedom mediates an 
attractive force, and modifies the gravity from the range 
determined by the Compton wavelength $\lambda=1/m$. 
From Eq.~(\ref{viablefr}), we have
\begin{eqnarray}
&&  f_{RR}={d^2f(R)\over dR^2}=4n(2n+1)\Lambda{R_c^{2n}\over R^{2n+2}}.
\end{eqnarray}

In the subhorizon limit, the matter density perturbation follows 
(e.g., \cite{Silvestri} and references therein), 
\begin{eqnarray}
\ddot \delta+2{\dot a\over a} \dot \delta -4\pi G_{\rm eff}(a,k) \rho \delta =0,
\label{deltaev}
\end{eqnarray}
where
\begin{eqnarray}
&&{G_{\rm eff}(a,k)\over G}=1+{1\over 3}{k^2/a^2\over k^2/a^2+1/(3f_{RR})},
\end{eqnarray}
and the dot denotes the differentiation with respect to the cosmic 
time. 

Instead of $R_c$, we introduce the parameter $k_c$ by 
\begin{eqnarray}
{1\over 3f_{RR}}=k_c^2
\left({{\Omega_0/a^3}+4({1-\Omega_0})\over \Omega_0+4({1-\Omega_0})}
\right)^{2n+2},
\end{eqnarray}
where $k_c$ represents the wavenumber corresponding to
the Compton wavelength at the present epoch. Thus,
the $f(R)$ model is specified by $n$ and $k_c$.
The growth factor can be obtained by solving Eq.~(\ref{deltaev}), 
which we denote by $D_1(a,k)$. 
The growth rate is given by $f=d\ln D_1(a,k)/d\ln a$. 

\begin{figure}
  \leavevmode
  \begin{center}
    \begin{tabular}{ c c }
      \includegraphics[width=2.6in,angle=0]{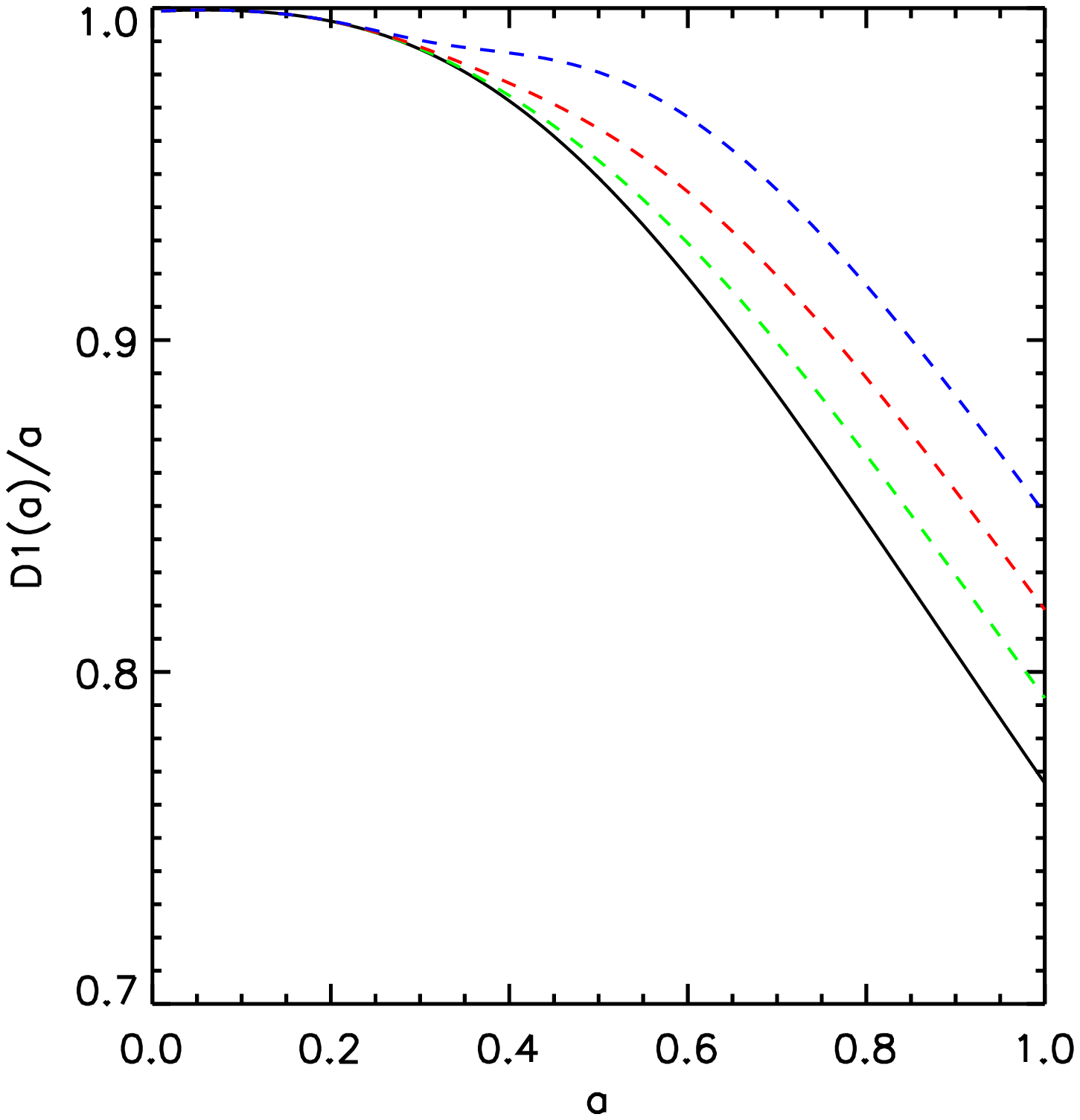}
      &
      \includegraphics[width=2.6in,angle=0]{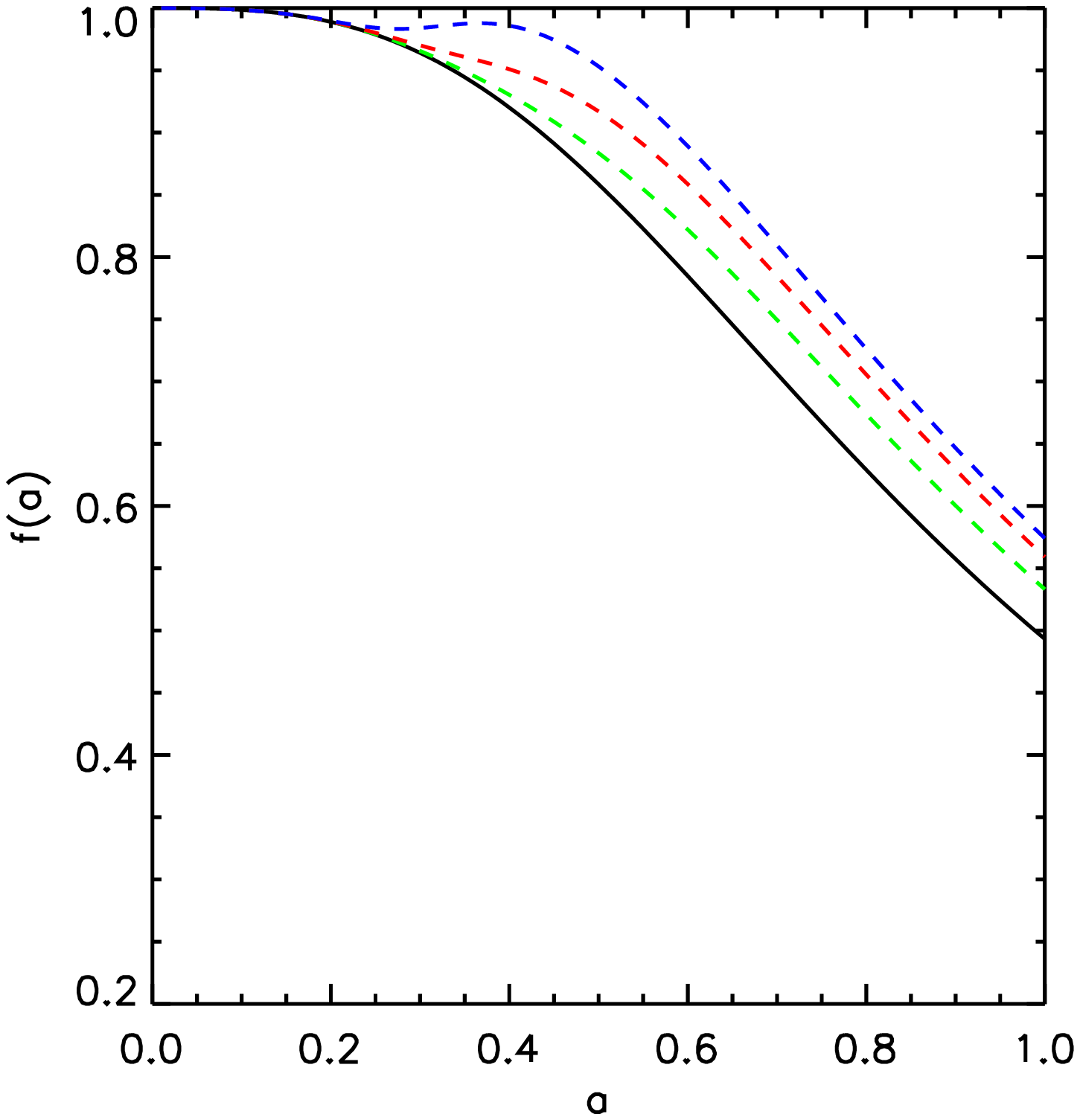}
    \end{tabular}
\caption{ (a,~Left) 
$D_1(a)/a$ as a function of the scale factor. 
The solid curve is the $\Lambda$CDM model with $\Omega_0=0.28$.
The dashed curves are for the $f(R)$ model with 
the wavenumbers $k/(h{\rm Mpc}^{-1})=0.2$, $0.1$, $0.05$ 
from the top to bottom, 
respectively. Here we adopted the model $n=1$ and $k_c=0.05h{\rm Mpc}^{-1}$. 
(b,~Right) Same as (a) but for the growth factor $f=d\ln D_1/ d\ln a$. 
}
\label{fig:modifiedgravityone}
\end{center}
\end{figure}

In the Einstein de Sitter background universe, the evolution 
of the density perturbation can be solved analytically \cite{Motohashi}. 
Two of the authors of the present paper investigated 
characteristic features of the evolution of the growth rate of 
the $f(R)$ model, both numerically and analytically 
in the reference \cite{Narikawa}. In the present paper, 
we solve the evolution equation (\ref{deltaev}) numerically
(cf. \cite{MotohashiII,MotohashiIII}). 
Figure \ref{fig:modifiedgravityone} shows the growth factor
divided by the scale factor (left) 
and the growth rate (right), respectively, as a
function of the scale factor.
The solid curve is the $\Lambda$CDM model with the density parameter 
$\Omega_0=0.28$. 
The dashed curves are for the $f(R)$ model with
different wavenumbers $k/(h{\rm Mpc}^{-1})=0.2,~0.1$,
and $0.05$, respectively. 
Here the $f(R)$ model assumes $n=1$ and $k_c=0.05h{\rm Mpc}^{-1}$. 
Due to the modification of the gravity the growth
factor and the growth rate are enhanced, and this enhancement is
scale-dependent.

\begin{figure}
  \begin{center}
    \includegraphics[width=0.4\textwidth]{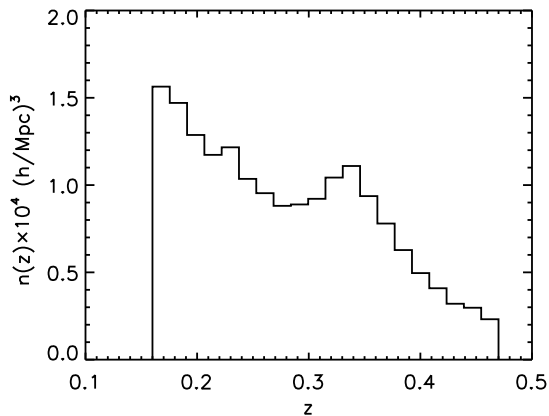}
     \end{center}
  \caption{ The mean number density of galaxies, $\bar n$, 
as a function of the redshift $z$ of the SDSS LRG sample, 
where we adopted the $\Lambda$CDM model with $\Omega_0=0.28$ 
for the distance-redshift relation $s=s[z]$.}
\label{fig:nz}
\end{figure}

\section{Multipole spectrum of the SDSS LRG sample}
The multipole power spectrum $P_\ell(k)$ is defined by the 
coefficient of the multipole expansion of the anisotropic
power spectrum $P(k,\mu)$, 
\begin{eqnarray}
  P(k,\mu)=\sum_{\ell=0,2,4\cdots}P_\ell(k) {\cal L}_\ell(\mu)(2\ell+1),
\end{eqnarray}
where ${\cal L}_\ell(\mu)$ are the Legendre polynomials,
$\mu(=\cos\theta)$ is the directional cosine between 
the line of sight direction and the wavenumber vector ${\bf k}$.
Note that our definition of the multipole spectrum 
$P_\ell(k)$ is different from the conventional one by the factor 
$2\ell+1$ \cite{Kaiser,Cole,Hamilton}.
Here the Legendre polynomials satisfy the normalisation condition,
\begin{eqnarray}
  \int_{-1}^{+1}d\mu{\cal L}_\ell(\mu){\cal L}_{\ell'}(\mu)
  ={2\over 2\ell+1}\delta_{\ell\ell'}.
\end{eqnarray}
The monopole $P_0(k)$ represents the angular averaged power spectrum, 
which is what we usually mean by the power spectrum; 
the quadrupole $P_{2}(k)$ represents the leading anisotropy in the power 
spectrum due to the redshift-space distortion. 
The hexadecapole $P_{4}(k)$ represents a different aspect of the 
redshift-space distortion. 
In the present paper, we focus on the monopole and quadrupole spectra. 
The quadrupole spectrum reflects the peculiar velocities of the
galaxies \cite{Kaiser,Cole,Hamilton}. Those peculiar motions can be used 
to test the gravity theory on cosmological scales. 

Pioneering works on the measurement of the quadrupole spectrum was
carried out by Cole, Fisher, and Weinberg \cite{Cole} and Hamilton 
\cite{Hamilton} using the IRAS galaxy survey catalogue. 
Cole et~al. presented a systematic method to estimate the quadrupole 
power spectrum through the anisotropic power spectrum \cite{Cole}. 
The method was applied to the Two Degree Field (2dF) galaxy survey
to estimate the $\beta$ factor. 
Hamilton obtained the quadrupole power spectrum by a transformation
of the correlation functions \cite{Hamilton}. 
In the present work, however, we adopt a different method to 
estimate the quadrupole power spectrum \cite{YNKBN}. 
Our method is in line to the widely used way to estimate the 
monopole power spectrum \cite{Feldman,Yamamoto2003}, and allows 
us to obtain the multipoles of the redshift-space power spectrum 
without evaluating the correlation function or the anisotropic 
power spectrum. In Ref.~\cite{Yamamoto08}, we applied the method to the
SDSS LRG sample from DR 6 to test the general relativity on cosmological 
scales. In the present paper, we revisit this problem with the SDSS LRG 
sample of DR 7 \cite{Abazajian}.

Our LRG sample is 
restricted to the redshift range $z=0.16-0.47$. 
In order to reduce the sidelobes of the survey 
window we remove some noncontiguous parts of 
the sample (e.g. three southern slices), which
leads us to $\sim 7150~{\rm deg}^2(={\Delta {\cal A}})$ sky 
coverage with a total of $N=100157$ LRGs. The data reduction 
procedure is the same as that described in 
\cite{Hutsi}. In this power spectrum analysis, we adopted
the spatially flat Lambda cold dark matter ($\Lambda$CDM) 
model distance-redshift relation $s=s[z]$, which is 
consistently chosen when comparing with theoretical prediction.

\begin{figure}
  \begin{center}
    \includegraphics[width=0.4\textwidth]{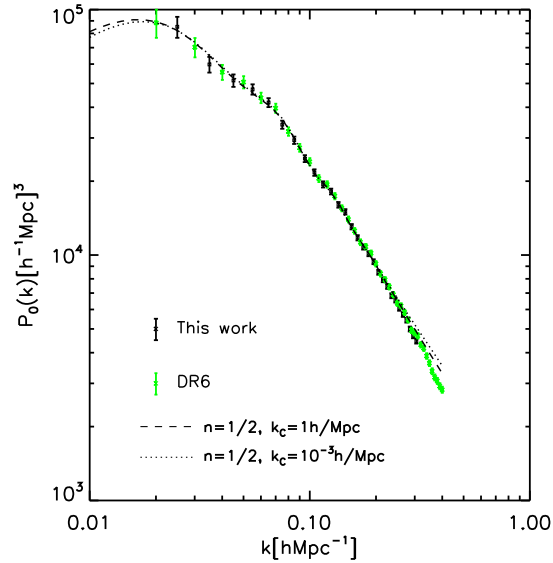}
     \end{center}
  \caption{ $P_0(k)$ of the SDSS LRG sample, where we adopted 
the distance-redshift relation $s=s[z]$ of the $\Lambda$CDM model
with $\Omega_0=0.28$. 
The dark (black) points correspond to the DR7, while the light (green) 
ones to the DR6.
The dashed and dotted curves show the $f(R)$
model with $n=1/2$, adopting a scale-dependent bias 
(case 1 in Eq.~(\ref{biascases})) and $\sigma_{\rm v}=350$km/s. 
The dashed curve is for $k_c=1h{\rm Mpc}^{-1}$, while the 
dotted curve is $k_c=10^{-3}h{\rm Mpc}^{-1}$. 
The cosmological parameters are
$\Omega_0=0.28$, $h=0.7$, and $n_s=0.96$ 
(primordial spectral index), and the amplitude of the 
perturbation is determined so as to be $\sigma_8=0.8$
in the limit of infinitely large $k_c$.
}
\label{fig:p0}
\end{figure}
\begin{figure}
  \begin{center}
    \includegraphics[width=0.4\textwidth]{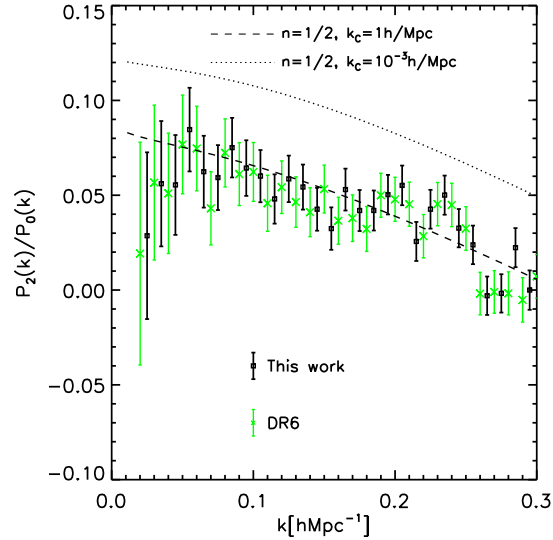}
     \end{center}
  \caption{ $P_2(k)/P_0(k)$ of the SDSS LRG sample. The meaning 
of the points corresponds to those of Fig.~\ref{fig:p0}. 
The dashed (dotted) curve is the theoretical prediction of the 
$f(R)$ model with $n=1/2$, $k_c=1h{\rm Mpc}^{-1}$ ($10^{-3}h{\rm Mpc}^{-1}$). 
The parameters of the bias model and $\sigma_{\rm v}$ are the same as 
those of Fig.~\ref{fig:p0}. 
The other cosmological parameters and the amplitude of the
primordial perturbation of the $f(R)$ model are also the same
as those of  Fig.~\ref{fig:p0}. }
\label{fig:p2}
\end{figure}

The strategy to measure the multipole power spectrum is the
same as that described in \cite{YNKBN}.  
We adopt the estimator of the multipole power spectrum 
for the discrete density field of the galaxy catalogue, as follows, 
\begin{eqnarray}
P_\ell(k)={1\over \Delta V_{k}}\int_{\Delta V_{k}} d^3 k
\left(R_{\ell}({\bf k})-S_\ell({\bf k})\right),
\label{estimatorPl}
\end{eqnarray}
where $\Delta V_{k}$ is the shell in the Fourier space and
\begin{eqnarray}
&&R_{\ell}({\bf k})=A^{-1}\left[
\sum_{i_1}^N\psi({\bf s}_{i_1},{\bf k})e^{i{\bfk}\cdot{\bf s}_{i_1}} 
{\cal L}_\ell(\hat {\bf s}_{i_1}\cdot \hat {\bf k})
-\alpha
\sum_{j_1}^{N_{\rm rnd}}\psi({\bf s}_{j_1},{\bf k})e^{i{\bfk}\cdot{\bf s}_{j_1}} 
{\cal L}_\ell(\hat {\bf s}_{j_1}\cdot \hat {\bf k})
\right]
\nonumber
\\
&&~~~~~~~~~~~~~~\times
\left[
\sum_{i_2}^N\psi({\bf s}_{i_2},{\bf k})e^{-i{\bfk}\cdot{\bf s}_{i_2}} 
-\alpha
\sum_{j_2}^{N_{\rm rnd}}\psi({\bf s}_{j_2},{\bf k})e^{i{\bfk}\cdot{\bf s}_{j_2}} 
\right],
\label{Rell}
\\
&&S_\ell({\bf k})=A^{-1}(1+\alpha) \sum_{i_1}^N
\psi({\bf s}_{i_1},{\bf k})
{\cal L}_\ell(\hat {\bf s}_{i_1}\cdot \hat {\bf k}),
\label{Sell}
\end{eqnarray}
where ${\bf s}_{i_1}$ (${\bf s}_{j_1}$) is the 
position of galaxies (random sample),
$\psi$ is the weight factor, which we take $\psi=1$, 
$\mu=\hat {\bf s}\cdot \hat {\bf k}$
is the directional cosine between $\hat{\bf s}(={\bf s}/|{\bf s}|)$ 
and $\hat {\bf k}(={\bf k}/|{\bf k}|)$, 
$\alpha\equiv N/N_{\rm rnd}$ in our case is $0.05$, and $A$ is determined by
\begin{eqnarray}
&&A=\int_{s(z_{\rm min})}^{s(z_{\rm max})} d{\bf s}\bar n^2(z) \psi^2({\bf s},{\bf k}).
\end{eqnarray}
Here the integral in the expression for $A$
means the integration over the whole survey volume, and $\bar n(z)$ is 
the mean (comoving) number density of the galaxies. 
The error for the estimator $P_\ell(k)$ is given 
by the variance \cite{YNKBN},
\begin{eqnarray}
\left<\Delta P_\ell(k)^2\right>\simeq2{(2\pi)^3\over \Delta V_k} {\cal Q}_l^2(k)
\label{DeltaPell}
\end{eqnarray}
with
\begin{eqnarray}
{\cal Q}_l^2(k)={1\over \Delta V_k} \int _{\Delta V_k} d{\bf k}
  A^{-2} \int_{s(z_{\rm min})}^{s(z_{\rm max})}d{\bf s} \bar n^4(z) \psi^4({\bf s},{\bf k})
[P({\bf k},{\bf s})+1/\bar n({\bf s})]^2
{\cal L}_\ell^2(\hat {\bf s}\cdot \hat {\bf k}).
\label{DeltaPellf}
\end{eqnarray}
Here we have assumed $\alpha\ll 1$. 
The covariance between the errors of different 
multipole spectra 
$\left<\Delta P_\ell(k)\Delta P_{\ell'}(k)\right>$ can be
evaluated with the same formulae
(\ref{DeltaPell}) and (\ref{DeltaPellf}), but only replacing 
${\cal L}_\ell^2(\hat {\bf s}\cdot \hat {\bf k})$ by  
${\cal L}_\ell(\hat {\bf s}\cdot \hat {\bf k})
{\cal L}_{\ell'}(\hat {\bf s}\cdot \hat {\bf k})$ in (\ref{DeltaPellf}) 
\footnote{
The contribution of the covariance between different multipoles 
was pointed out by Atsushi Taruya and Masahiro Takada.}.
In our analysis we adopt 
$\psi({\bf s},{\bf k})=1$. Figure \ref{fig:nz} shows the
mean number density as a function of $z$, when assuming
the $\Lambda$CDM with $\Omega_0=0.28$ for the distance-redshift
relation $s=s[z]$.

Figure \ref{fig:p0} compares the observed monopole power spectrum 
and our theoretical model.
The dark(black) points with error bars in figure \ref{fig:p0} show
the monopole power spectrum of the DR7. The 
light(green) points are the previous results for 
the DR6 \cite{Yamamoto08}. The dashed and the dotted curves
represent the $f(R)$ model with $n=1/2$,
with the scale-dependent bias model of case 1 (see the next section 
for details). 
The dashed curve is for $k_c=1h{\rm Mpc}^{-1}$, while the 
dotted one for $10^{-3}h{\rm Mpc}^{-1}$. 
The cosmological parameters are
$\Omega_0=0.28$, $h=0.7$, $n_s=0.96$ (primordial spectral index).
The amplitude of the primordial perturbation is chosen to 
be $\sigma_8=0.8$ in the limit of infinitely large $k_c$.  
The Smith's nonlinear fitting formula \cite{SM} is adopted. One can 
see that $P_0(k)$ can be fitted with our theoretical model,
by choosing suitable bias parameters.

Figure \ref{fig:p2} plots $P_2(k)/P_0(k)$. The meaning of the
points and the parameters of the curves corresponds to those 
of Fig.~\ref{fig:p0}.
This figure shows that the quadrupole power spectrum can 
be used to constrain the $f(R)$ model. 
Also it is clear that the long Compton wavelength model doesn't 
fit the data. 


\begin{figure}
\leavevmode
\begin{center}
\begin{tabular} {c}
    \includegraphics[width=0.8\textwidth]{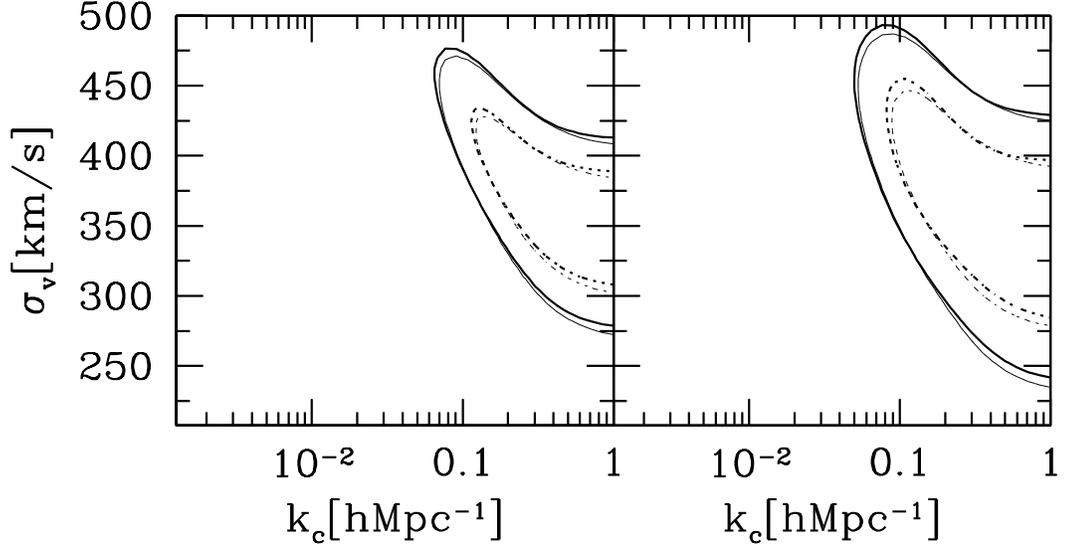}
\end{tabular}
\end{center}
  \caption{$\Delta \chi^2$ on the $k_c-\sigma_{\rm v}$ plane. Here
we adopted the model $n=1/2$. The other parameters are
$\Omega_0=0.28$, $h=0.7$, $n_s=0.96$. The normalisation is 
fixed $\sigma_8=0.8$ in the limit of large $k_c$.
The Peacock and Dodds's nonlinear fitting formula is used
for the thin curves, while the Smith formula is used 
for the thick curves. Solid (dashed) contours correspond to
$\Delta \chi^2=6.2$ ($2.3$).
The left panel adopted Eq.~(\ref{covarinceFKP}), while 
the right panel the covariance matrix from the mock catalogues. 
}
\label{fig:kcsigmav}
\end{figure}

\begin{figure}
\leavevmode
\begin{center}
    \begin{tabular}{ l r }
    \includegraphics[width=0.43\textwidth]{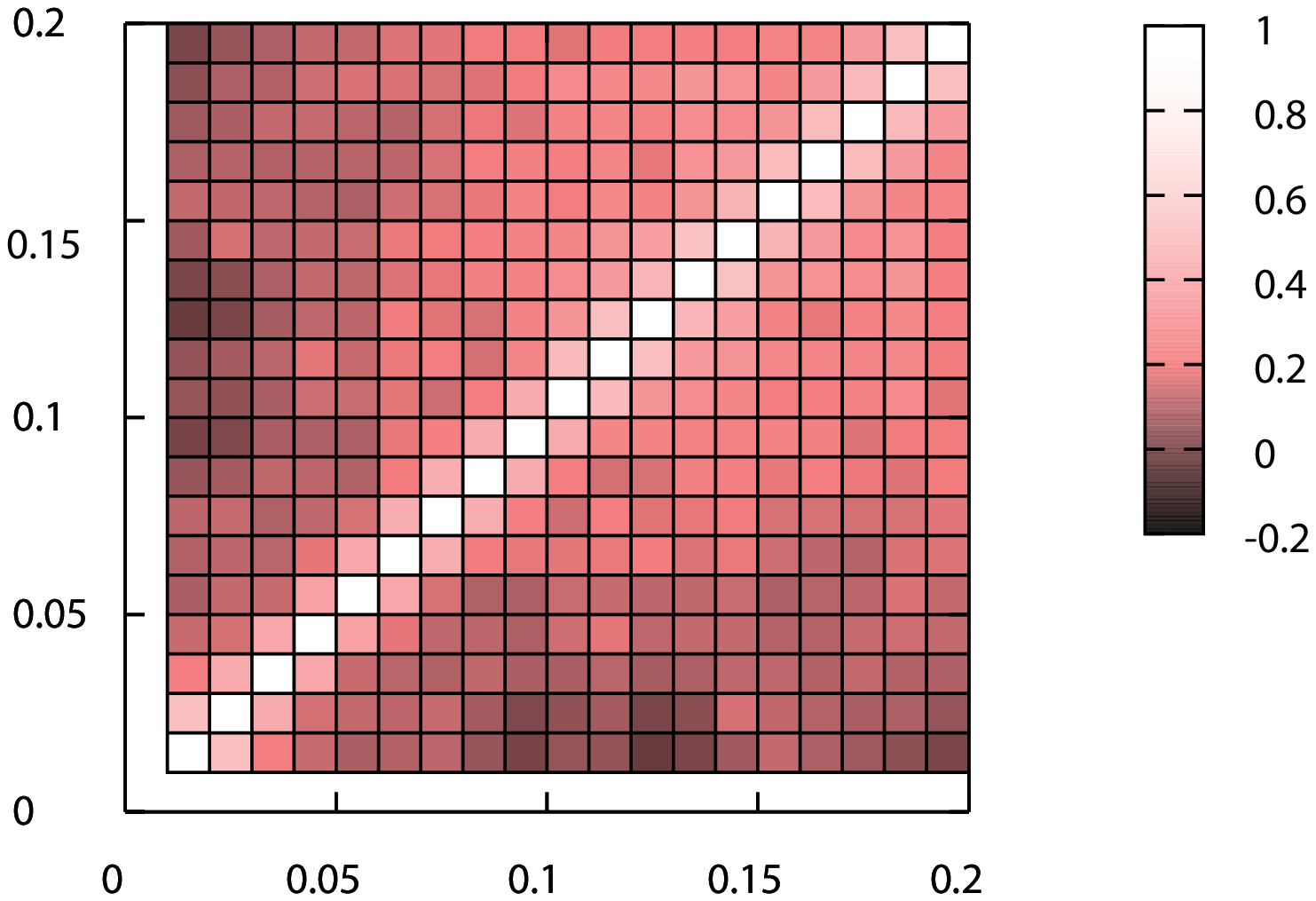}
&
    \includegraphics[width=0.43\textwidth]{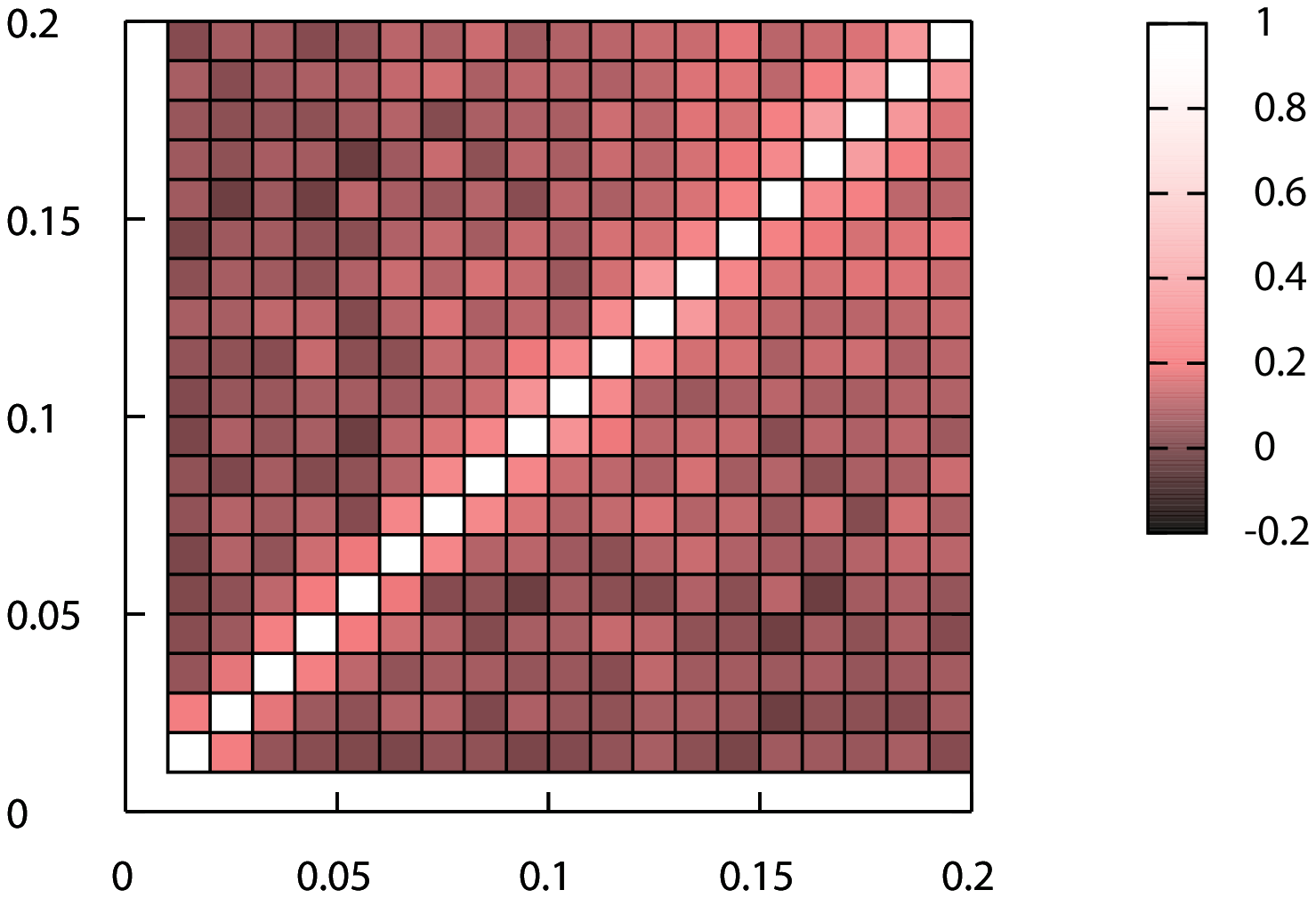}
\end{tabular}
\end{center}
\caption{The correlation matrix, Eq.~(\ref{correlationmatrix}), 
for $\ell=0$ (left) and $\ell=2$ (right), respectively, 
from $1000$ mock catalogues. }
\label{fig:correlationmatrix}
\end{figure}

\begin{figure}
\leavevmode
\begin{center}
\begin{tabular} {c}
    \includegraphics[width=0.57\textwidth]{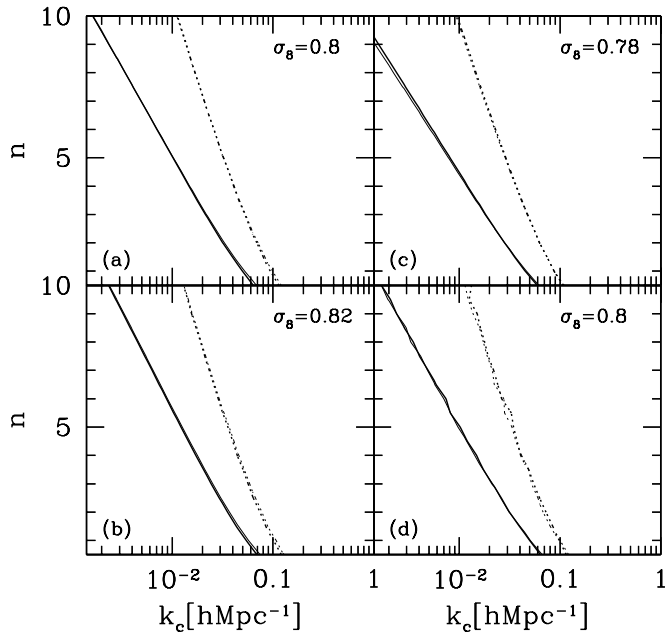}
\end{tabular}
\end{center}
\caption{
$\Delta\chi^2$ on the $k_c-n$ plane, which we 
evaluated with Eqs.~(\ref{chisquarecorr}) and 
(\ref{covarinceFKP}).
For each pair of $k_c$ and $n$, the minimum
value of $\chi^2$ is computed by fitting 
the bias parameter and $\sigma_{\rm v}$.
Other parameters are fixed 
$\Omega_0=0.28$, $h=0.7$, and $n_s=0.96$.
The normalisation of the primordial perturbation is
chosen so as to be $\sigma_8=0.8$ (a)
$\sigma_8=0.82$ (b), and $\sigma_8=0.78$ (c), 
in the limit of large $k_c$.
The panels (a)-(c) adopt the bias model of case 1. 
The panel (d) is the same as (a) but with 
bias model of case 2.
Solid (dotted) contours correspond to $\Delta \chi^2=6.2$ ($2.3$).
Almost overlapping thin and thick curves assume the 
Peacock and Dodds's formula and the Smith's formula, respectively. 
}
\label{fig:kcnPDSM}
\end{figure}
\begin{figure}
\leavevmode
\begin{center}
\begin{tabular} {c}
    \includegraphics[width=0.57\textwidth]{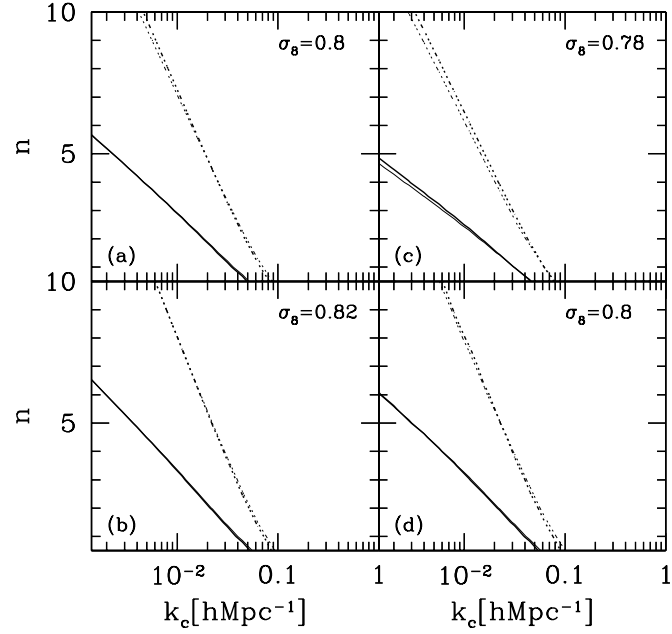}
\end{tabular}
\end{center}
\caption{The same as the Fig.~\ref{fig:kcnPDSM} but with 
the covariance matrix from the mock catalogues.}
\label{fig:kcnPDSMcorr}
\end{figure}
\begin{figure}
\leavevmode
\begin{center}
\begin{tabular} {c}
    \includegraphics[width=0.50\textwidth]{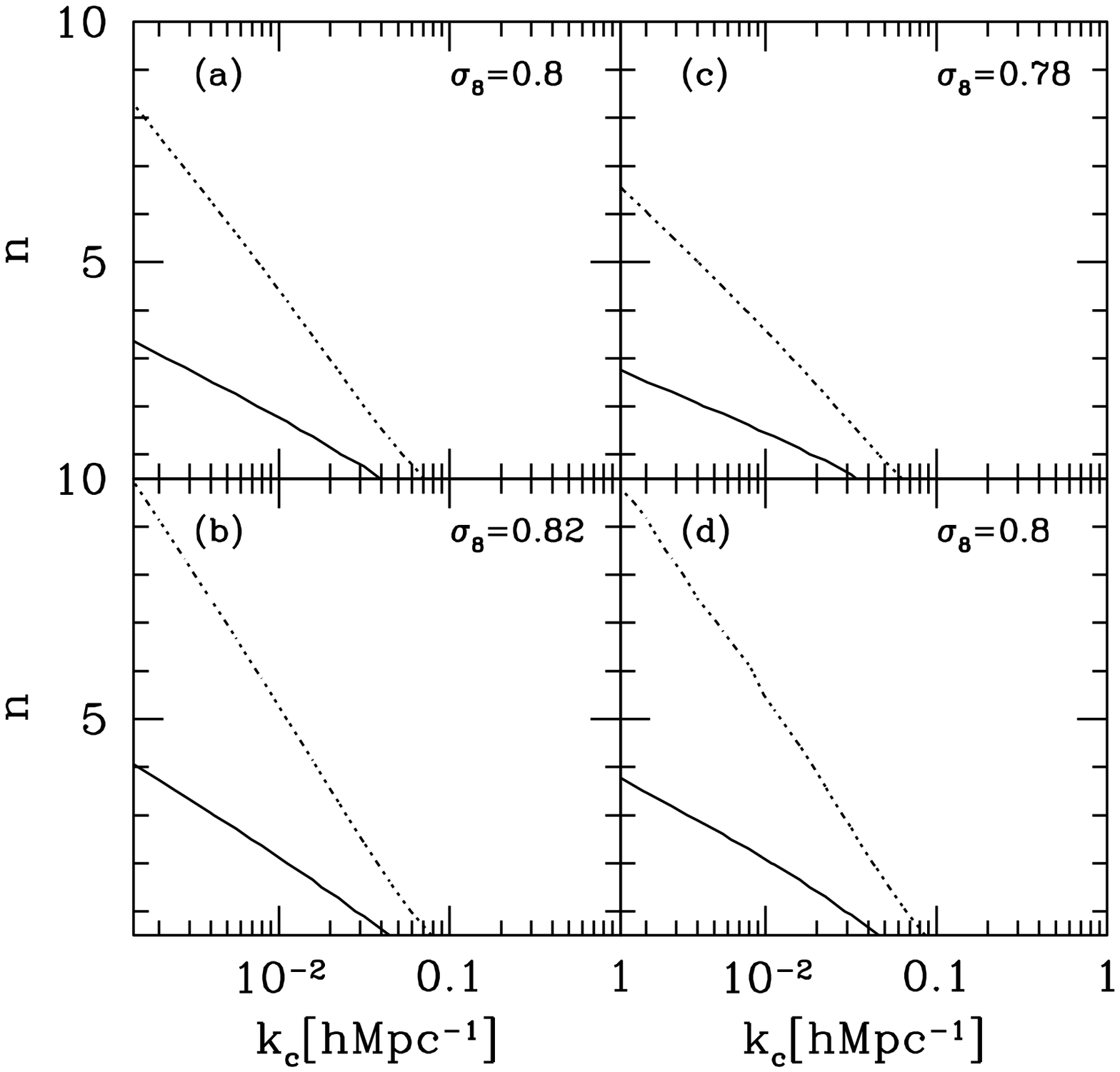}
\end{tabular}
\end{center}
\caption{The same as the Fig.~\ref{fig:kcnPDSM} but with 
the covariance matrix from the mock catalogues and the 
redshift-space power spectrum (\ref{JE}). 
Only the curves with Peacock Dodds's formula for the nonlinear matter
power spectrum are plotted. }
\label{fig:kcnPDSMcorrJE}
\end{figure}

\section{Cosmological constraint}
\def\wsigmav{{{\widetilde \sigma}_{\rm v}}}
In order to investigate the cosmological constraint on the 
$f(R)$ model from the multipole spectra, our theoretical model
needs to include nonlinear effects.
In the present paper, for simplicity, we adopt the following model 
for the galaxy power spectrum \cite{PD94,CFW}, 
\begin{eqnarray}
P_{\rm gal} 
  \left(k,\mu,z\right)=(b+f\mu^2)^2P_{\rm nl}(k,z)
{\cal D} \left[{\sigma_{\rm v}k\mu}\right],
\label{Pgalaxyn}
\end{eqnarray}
where $P_{\rm nl}(k,z)$ denotes a nonlinear matter power 
spectrum, ${\cal D}[k\mu{\sigma_{\rm v}}]$ is the damping 
factor due to the Finger of God effect, and $\sigma_{\rm v}^2$ 
is the pairwise velocity dispersion.
Assuming an exponential distribution function for the 
pairwise velocity, $e^{-\sqrt{2}|v_{12}|/\sigma_{\rm v}}/\sqrt{2}\sigma_{\rm v}$, 
where $v_{12}$ is the pairwise peculiar velocity projected along
the separation of a pair, the damping function is \cite{FOG} 
(cf. \cite{Peebles76,Park}),
\begin{eqnarray}
  {\cal D} \left[{\sigma_{\rm v}k\mu}\right]=
  {1\over 1+{\wsigmav^2 k^2\mu^2/ 2}}
\label{defcalD}
\end{eqnarray}
with $\wsigmav=\sigma_{\rm v}/H_0$. In this case, we have 
\begin{eqnarray}
&&P_0(k,z)={1\over 3k^5 \wsigmav^5}\left[
2fk\wsigmav(-6f+(6b+f)k^2\wsigmav^2)+3\sqrt{2}(-2f+bk^2\wsigmav^2)^2
{\rm tan}^{-1}{k\wsigmav\over \sqrt{2}}
\right]P_{\rm nl}(k,z),
\\
&&P_2(k,z)={1\over 30k^7 \wsigmav^7}\biggl[
-360 bf k^3\wsigmav^3+90b^2 k^5 \wsigmav^5+8f^2k\wsigmav(45+k^4\wsigmav^4)
\nonumber
\\
&&\hspace{3cm}-15\sqrt{2}(6+k^2\wsigmav^2)(-2f+bk^2\wsigmav^2)^2{\rm tan}^{-1}{k\wsigmav\over \sqrt{2}}
\biggr]P_{\rm nl}(k,z),
\\
&&P_4(k,z)={(-2f+bk^2\wsigmav^2)^2\over 24k^9 \wsigmav^9}\left[
-10k\wsigmav(42+11k^2\wsigmav^2)+3\sqrt{2}(140+60k^2\wsigmav^2+3k^4\wsigmav^4)
{\rm tan}^{-1}{k\wsigmav\over \sqrt{2}}
\right]P_{\rm nl}(k,z),
\end{eqnarray}
from Eqs.~(\ref{Pgalaxyn}) and (\ref{defcalD}).
For the nonlinear matter power spectrum, $P_{\rm nl}(k,z)$, we adopt the
fitting formulas by Peacock and Dodds \cite{PD96} or by Smith et al. 
\cite{SM}.
For the bias, we consider the following scale-dependent forms,
\begin{eqnarray}
b(k)=\left\{
\begin{array}{ll}
\displaystyle{b_0+b_1\left({k\over 0.1h{\rm Mpc}^{-1}}\right)^\alpha}
&({\rm case~1})
\\
\displaystyle{b_0+b_1\left({k\over 0.1h{\rm Mpc}^{-1}}\right)+
b_2\left({k\over 0.1 h{\rm Mpc}^{-1}}\right)^2}
&({\rm case~2})
\end{array}
\right. ,
\label{biascases}
\end{eqnarray}
where $b_0,~b_1,~b_2$, and $\alpha$ are the fitting parameters. 

Our strategy is the following. We use the monopole and 
quadrupole spectra in the wavenumber range 
$0.02 h{\rm Mpc}^{-1}\leq k_i\leq 0.2 h{\rm Mpc}^{-1}$, 
and compute the chi squared
\begin{eqnarray}
 \chi^2=\sum_{\ell,\ell'=0,2}\sum_{i,j}(P_\ell(k_i)-P_\ell^{\rm obs}(k_i))
{C}_{\ell\ell'}^{-1}(k_i,k_j)(P_{\ell'}(k_j)-P_{\ell'}^{\rm obs}(k_j)),
\label{chisquarecorr}
\end{eqnarray}
where $P_\ell^{\rm obs}(k_i)$ is the observed power spectrum and 
${C}_{\ell\ell'}(k_i,k_j)=\left<\Delta P_\ell(k_i)\Delta P_{\ell'}(k_j)\right>$
is the covariance matrix.
Here the covariance of the errors of the monopole and 
quadrupole spectra is taken into account, however it does not 
affect our results quantitatively.

The left panel of Fig.~\ref{fig:kcsigmav} shows the contours of 
$\Delta \chi^2$ on the $k_c-\sigma_{\rm v}$ plane,
where we used the covariance matrix from section 3,
\begin{eqnarray}
{C}_{\ell\ell'}(k_i,k_j)
=\left<\Delta P_\ell(k_i)\Delta P_{\ell'}(k_j)\right>\delta_{ij}.
\label{covarinceFKP}
\end{eqnarray}
The one-sigma (dashed curve) and two-sigma (solid curve) contour-levels are
given, respectively.  
Here the chi squared is computed to minimise (\ref{chisquarecorr})
by fitting the bias parameters $b_0,~b_1,~b_2$, or $\alpha$,
for each value of $k_c$ and $\sigma_{\rm v}$. 
The other parameters are fixed $n=1/2$, $\Omega_0=0.28$, 
$\Omega_b=0.044$, $n_s=0.96$, and $h=0.7$. 
For $P_{\rm nl}(k,z)$, we adopted the Peacock and Dodds's 
formula \cite{PD96} (thin curve) and the Smith formula \cite{SM} 
(thick curve), respectively. 
The redshift is fixed to $z=0.3$, which is typical for the LRG sample. 
The amplitude of the matter power spectrum is fixed so as
to be $\sigma_8=0.8$ in the limit of infinitely large $k_c$, 
i.e., in the limit of the $\Lambda$CDM model. 

For comparison, the right panel of Fig.~\ref{fig:kcsigmav} 
shows the contours of $\Delta \chi^2$, which take the correlation 
of the errors of different wavenumbers 
into account by evaluating Eq.~(\ref{chisquarecorr}),
with 
the covariance matrix obtained from mock catalogues. 
Due to the inclusion of the correlation 
of errors of different wavenumbers, the constraint becomes weaker 
compared with the left panel.

In the right panel of Fig.~\ref{fig:kcsigmav}, 
we obtain the covariance matrix by using 
mock catalogues, which were built by following the procedure 
described in the reference \cite{Hutsi}. 
First, we generate density field using a second order Lagrangian 
perturbation calculation. Then, we perform Poisson sampling of 
the generated density field so as to end up with a galaxy sample
that has a clustering strength enhanced by a bias and a number 
density equal to the observed LRG sample density. 
We then extract the catalogue by applying the radial and angular
selection function. 
We have checked that the mock catalogues have the  
amplitude of the monopole and quadrupole power spectra
consistent with the observed LRG power spectra,  
and also that the diagonal components of the 
covariance matrix from the mock catalogues give almost the same error 
as those of Eq.~(\ref{DeltaPell}) in the range of
$0.02 h{\rm Mpc}^{-1}\leq k_i\leq 0.2 h{\rm Mpc}^{-1}$ \cite{Hutsi,sato}.
Figure \ref{fig:correlationmatrix} shows the two dimensional map 
of the correlation matrix, 
\begin{eqnarray}
r_\ell(k_i,k_j)={C_{\ell\ell}(k_i,k_j)\over \sqrt {C_{\ell\ell}(k_i,k_i)
C_{\ell\ell}(k_j,k_j)}},
\label{correlationmatrix}
\end{eqnarray}
for $\ell=0$ and $2$ from $1000$ mock catalogues.
The binning of the covariance matrix is $\Delta k=0.01~h{\rm Mpc}^{-1}$. 
One can see from Fig.~\ref{fig:correlationmatrix} 
that the off diagonal part is suppressed.

The normalisation of the cosmological perturbations should be 
determined by the cosmic microwave background anisotropies, 
depending on the parameters $n$ an $k_c$ of the $f(R)$ model. 
However, the background expansion of the viable $f(R)$ model 
is almost the same as that of the $\Lambda$CDM model, and
the evolution of the matter density perturbations 
is only altered at late time, if compared with the $\Lambda$CDM model.
This alteration will raise an additional integrated Sachs Wolfe effect 
on the CMB anisotropies due to the modified evolution of the matter 
density perturbation at late time. 
We neglect this effect on the normalisation of the perturbation, 
for simplicity. 
Then, we simply fixed the amplitude of the primordial cosmological 
perturbation by $\sigma_8$ in the limit of large $k_c$, 
i.e., the $\sigma_8$ of the $\Lambda$CDM model. 

Figure \ref{fig:kcsigmav} shows that the shorter Compton wavelength 
model with $\sigma_{\rm v}\simeq350{\rm km/s}$ gives the best fit to 
the data.
Figure \ref{fig:kcnPDSM} shows the contours of 
$\Delta \chi^2$ on the $k_c-n$ plane. Here $\chi^2$ is
computed with Eq.~(\ref{chisquarecorr}) with (\ref{covarinceFKP}) 
by fitting the bias parameters and $\sigma_{\rm v}$. 
The panels (a), (b), and (c) fix the normalisation of the 
perturbation to be $\sigma_8=0.8$, $0.82$ and $0.78$, 
in the limit of large $k_c$, respectively.  
The contour levels of $\Delta\chi^2=2.3$ (dotted curve)
and $6.2$ (solid curve), correspond to $1\sigma$ and 
$2\sigma$ confidence, respectively.
In figure~\ref{fig:kcnPDSM} we used the Peacock and Dodds's formula 
(thin curve) and the Smith formula \cite{SM} 
(thick curve), respectively, though the two curves almost overlap. 
The panels (a), (b), and (c) adopt the bias model of case 1. 
The panel (d) is the same as (a), but adopted the bias model 
of case 2. The left lower region in each panel is excluded. 

Figure \ref{fig:kcnPDSMcorr} is the same as Fig.~\ref{fig:kcnPDSM}, but
adopted the covariance matrix from the mock catalogues for the chi squared.
The constraint of Fig.~\ref{fig:kcnPDSMcorr} is weaker compared with 
that of Fig.~\ref{fig:kcnPDSM}.
Especially, the constraint for the model with larger $n$ 
becomes weaker. However, 
Fig.~\ref{fig:kcnPDSMcorr} indicates that the long Compton 
wavelength case of the $f(R)$ model with the smaller value 
of $n$ is excluded.

Thus far, we have used the redshift-space power spectrum (\ref{Pgalaxyn}). 
In order to check the reliability of our result, we next consider 
the other possible model for the redshift-space power spectrum,
\begin{eqnarray}
P_{\rm g}(k,\mu)=\left(
b^2(k)P_{\delta\delta}(k)+2fb(k)P_{\delta\theta}(k)\mu^2
+f^2P_{\theta\theta}(k)\mu^4 \right)e^{-(fk\mu\sigma_v)^2},
\label{JE}
\end{eqnarray}
where $P_{\delta\delta}(k)$ is the nonlinear matter power spectrum,
$P_{\theta\theta}(k)$ is the power spectrum of the velocity 
divergence, and $P_{\delta\theta}(k)$ is the cross power 
spectrum of matter and the velocity divergence. 
This model is obtained from the model proposed by Scoccimarro 
\cite{Scoccimarro} and assumes a linear bias relation.
Very recently, Jenning et al. proposed a fitting formula for
the redshift-space power spectrum of the form (\ref{JE}), assuming 
$b(k)=1$. The fitting formula relates the nonlinear
matter power spectrum $P_{\delta\delta}(k)$ to $P_{\delta\theta}(k)$ 
and $P_{\theta\theta}(k)$. 
By using the N-body simulations it was demonstrated that the fitting 
formula is accurate to better than  $10$\% for the $\Lambda$CDM model 
and quintessence dark energy models for $k\simlt 0.2h{\rm Mpc}^{-1}$.
Although the accuracy of the fitting formula for the $f(R)$ model 
has not been explicitly demonstrated, we assume its validity,
and use it in the following $\Delta \chi^2$ calculations.  

Figure \ref{fig:kcnPDSMcorrJE} shows the contours of 
$\Delta \chi^2$ on the $k_c-n$ plane, the same as 
Fig.~\ref{fig:kcnPDSM}, but with 
the covariance matrix from the mock catalogues and the 
redshift space power spectrum (\ref{JE}). 
In the original formula, $\sigma_v$ is obtained from 
$P_{\theta\theta}(k)$, however, we assumed $\sigma_v$
to be a fitting parameter, as is done in Fig.~\ref{fig:kcnPDSMcorr}.
This figure shows that the constraint becomes weaker when compared
to the previous model (\ref{Pgalaxyn}). 
The models with large value of $n$ are not constrained. 
However, the long Compton wavelength case of the $f(R)$ 
model with the smaller value of $n$ is excluded.
This new model predicts that $P_{\delta\theta}(k)$ is smaller
than $P_{\delta\delta}(k)$ for values of 
$k\simlt0.1{\rm Mpc}^{-1}$, which reduces the quadrupole 
power spectrum and thus weakens the constraint. 

Let us compare our result with the other
constraints on the $f(R)$ model.  
Refs.~\cite{SPH,Schmidt} have investigated the constraints on 
the $f(R)$ model for the case $n=1/2$.
In Ref.~\cite{SPH}, the constraint from the CMB anisotropies
through the integrated Sachs Wolfe effect is investigated. 
However, the constraint is weak. Only the horizon-scale 
Compton wavelength model is excluded. 
In Ref.~\cite{Schmidt}, the constraint from the cluster 
number count is investigated. Though it is restricted to 
the case $n=1/2$, they obtained $|f_{R0}|\simlt 10^{-4}$, 
where $f_{R0}$ is the value of $f_{R}$ at the present epoch.
In the case $n=1/2$, $|f_{R0}|$ is related to $k_c$ by
\begin{eqnarray}
k_c\simeq0.04\left({10^{-4}\over |f_{R0}|}\right)^{1/2}h{\rm Mpc}^{-1}.
\end{eqnarray}
Ref.~\cite{Schmidt} reports that 
$k_c\simlt0.04h{\rm Mpc}^{-1}$ is excluded. 
The constraint is similar to our result, when the redshift-space 
power spectrum (\ref{Pgalaxyn}) is used (See Figure
\ref{fig:kcnPDSMcorr}).
When arguably more accurate model (\ref{JE}) is used,
the constraint becomes slightly weaker than that of (\ref{Pgalaxyn})
(See Figure \ref{fig:kcnPDSMcorrJE}).

\section{Future prospect of measuring Compton scale}
In this section, we estimate future prospects of constraining
the Compton scale with the use of the Fisher matrix technique, 
which is frequently used for estimating minimal attainable
constraint on model parameters. We focus on the 
error of the Compton wavenumber $k_c$. We adopt the
Fisher matrix of the form (e.g., \cite{Verde}), 
\begin{eqnarray}
F_{ij}= {1\over 4\pi^2}\int_{k_{\rm min}}^{k_{\rm max}} dk k^2
\int_{-1}^{+1}d\mu {\partial P_{\rm gal}(k,\mu)\over \partial \theta^i}
                 {\partial P_{\rm gal}(k,\mu)\over \partial \theta^j}
                {V\over (P_{\rm gal}(k,\mu)+1/\bar n)^2},
\end{eqnarray}
where $\theta^i$ denotes a model parameter, $V$ is a survey volume, 
$\bar n$ is a mean number density of galaxies. 

\begin{figure}
  \leavevmode
  \begin{center}
      \includegraphics[width=0.6\textwidth]{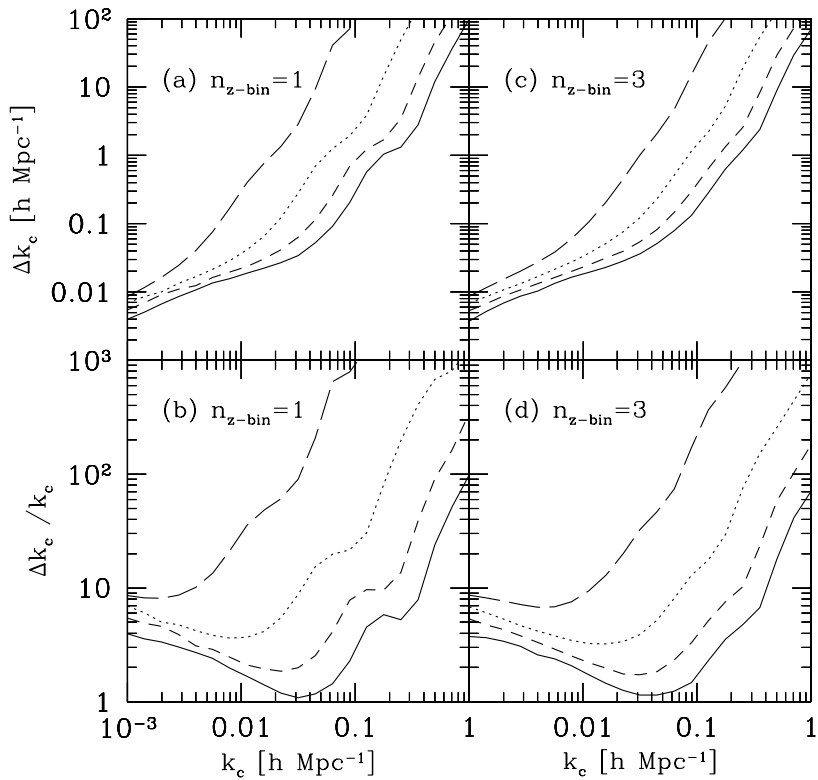}
     \end{center}
  \caption{(a) $1\sigma$ error $\Delta k_c$ as a function
of the target value of $k_c$. The result is based on the
Fisher matrix analysis with the $6$ parameters, 
$k_c$, $n$, $\sigma_{\rm v}$, and $b_0$, $b_1$ and $\alpha$ for 
the bias model 1, and marginalized over the $5$ parameters
other than $k_c$. 
The target parameters are $b_0=2.5$, $b_1=0.5$,  $\alpha=1/2$, 
and $n$ is chosen $n=1/2$, $1$, $2$, $4$ from the bottom to the top,
respectively. 
The other parameters are fixed
$\Omega_0=0.28$, $h=0.7$, $n_s=0.96$, and the normalisation 
$\sigma_8=0.8$ in the limit of the $\Lambda$CDM model.
Eq.~(\ref{Pgalaxyn}) with the Peacock and Dodds nonlinear fitting 
formula is adopted. 
(b) the relative error $\Delta k_c/k_c$ corresponding to (a).
(c) and (d) are the same as (a) and (b), respectively,  
but assumed the analysis where the full sample is divided into 3 
redshift bins.}
\label{fig:sumire}
\end{figure}

In the Fisher matrix analysis,
for simplicity, we consider the $6$ parameters 
$k_c$, $n$, $\sigma_{\rm v}$, $b_0$, $b_1$ and 
$\alpha$, adopting the bias model of case 1.
The panel (a) of Fig. \ref{fig:sumire} shows the $1\sigma$ error 
$\Delta k_c$, in determining the Compton wavenumber 
$k_c$ as a function of the target value of $k_c$, 
assuming a redshift survey like the SUMIRE (SUbaru Measurement 
of Imaging and REdshift of the universe) \cite{sumire}, 
which assumes the survey parameters like those of the WFMOS 
survey \cite{WFMOS}, the range of the redshift $0.9<z<1.6$, the 
survey area $2000$ square degrees,
and the mean number density $\bar n=4\times 10^{-4}(h^{-1}{\rm Mpc})^{-3}$.
Here we adopted the target values $\sigma_{\rm v}=400$km/s, 
$b_0=2.5$, $b_1=0.5$, $\alpha=0.5$, and $n=1/2$, $1$, $2$, 
and $4$, from the bottom to the top, respectively.
The other parameters are fixed $\Omega_0=0.28$, $h=0.7$, $n_s=0.96$, 
and the normalisation so as to be $\sigma_8=0.8$ in the limit of 
the $\Lambda$CDM model.
We obtained $\Delta k_c$ by marginalizing 
the Fisher matrix over the $5$ parameters 
$n$, $\sigma_{\rm v}$, $b_0$, $b_1$ and $\alpha$.
The panel (b) of Fig. \ref{fig:sumire} shows the relative 
error $\Delta k_c/k_c$. 

In the Fisher matrix we used the power spectrum in the range
of wavenumbers $k<0.3h{\rm Mpc}^{-1}$. 
This immediately implies that the redshift survey cannot be very sensitive
to the models with the short Compton wavelength, as seen from 
figure \ref{fig:sumire}.
The error becomes very 
large for $k_c\simgt0.1h{\rm Mpc}^{-1}$, but it will be possible
to obtain a useful constraint on the Compton scale, in principle, 
for models with $k_c\simlt0.1h{\rm Mpc}^{-1}$.
However, the constraint becomes weak for the case of large $n$. 

The panel (a) assumes the power spectrum analysis without dividing the 
full galaxy sample, which spans the redshift range $0.9 \leq z \leq 1.6$, into 
redshift subsamples.
The panel (c) assumes the case when 
the galaxy sample is divided into the three subsample in 
redshift bins and that the power spectra are obtained from each subsample. 
In this case, the parameters $\sigma_{\rm v}$, $b_0$, $b_1$ and 
$\alpha$ should be fitted in each redshift bin, and the total 
number of parameters in the Fisher matrix analysis is 14.
The panel (d) is the relative error, corresponding to (c). 
The cosmological parameters are the same as those of (a). 
The possible advantage of this method is that the additional
information of the redshift evolution might improve the
constraint. One can see that the constraint is improved in comparison
with the panel (a) or (b). The degree of the improvement is 
small for $n=1/2$, but is not negligible for the case $n=4$. 
This is understood because the redshift evolution of the Compton 
scale is faster for larger $n$.

\section{Summary and conclusions}
In this paper, we determined a cosmological constraint
on the viable $f(R)$ model based on the redshift-space distortion
by measuring the monopole and quadrupole spectra 
of the SDSS LRG sample of DR7. 
The monopole and the quadrupole spectra are used to fit the
bias parameters and to constrain the growth factor and the growth 
rate of the density perturbations, which depend on the Compton
scale of the $f(R)$ model.

Our results show that short Compton wavelength model 
fits the data better, while the long Compton wavelength model is
excluded, though the constraint depends on the evolution 
parameter $n$. For the case $n=1/2$, our constraint is 
similar to that from the cluster number counts reported in 
\cite{Schmidt}. 
When we adopt more accurate model for the 
redshift-space power spectrum \cite{Jennings}, the
constraint becomes slightly weaker. However, the long Compton 
wavelength case of the $f(R)$ 
model with the smaller value of $n$ is excluded. 
Our results exemplify that the redshift-space distortion 
is quite useful in testing gravity theory. 
We also demonstrated that a future redshift survey like 
the WFMOS/SUMIRE is potentially useful in obtaining a constraint 
on the Compton wavelength scale.

We acknowledge that the widely used theoretical model 
of the anisotropic power spectrum adopted in the present
paper might need careful improvements.
We adopted the Peacock and Dodds formula and the Smith
formula for the nonlinear modelling of the mass power spectrum. 
Our results do not significantly depend on the choice. 
However, there might be a need to adopt a more sophisticated formula 
for the precise nonlinear modelling within the framework of the modified 
gravity, as demonstrated by Koyama, Taruya, Hiramatsu 
\cite{KTH}. 
The treatment of the Finger of God effect in our paper was simple, 
which assumed the exponential distribution function for the pairwise 
velocity and introduced one free parameter -- the pairwise velocity 
dispersion. In reality it might not be an adequate model to describe the 
nonlinear region of the redshift-space power spectrum~\cite{Scoccimarro}. 
We checked the reliability of our results by adopting the other possible 
model proposed in Ref.\cite{Jennings}, extensively
applying the fitting formula to the $f(R)$ model, 
whose accuracy in this case, however, has not been demonstrated. 
We found that there is a
non-negligible effect on the constraint on the $f(R)$ model. 
Therefore, a more precise modelling of the redshift-space power 
spectrum should arguably be needed in the future.  
Concerning the modelling of the clustering bias, we adopted 
a simple scale-dependent bias. Here too there is potentially a lot of room for 
improvement. These issues are out of scope 
for the present paper, but need to be elaborated for a precise test 
of gravity with the future redshift surveys.


\vspace{0.3cm}
{\it Acknowledgement}
We thank T.~Kobayashi, J.~Yokoyama, H.~Motohashi, T.~Nishimichi, 
S.~Saito, A.~Taruya and M.~Takada, 
and Y.~Suto for useful comments and discussions.
This work was supported by Japan Society for Promotion
of Science (JSPS) Grants-in-Aid
for Scientific Research (Nos.~21540270,~21244033).
This work is also supported by JSPS 
Core-to-Core Program ``International Research 
Network for Dark Energy''.
T.N. acknowledges support by a research assistant program
of Hiroshima University.

\vspace{0.3cm}

\end{document}